# Valorisation of Fermentation Side-Stream for Waste-to-Mycoprotein: Nutrient Composition, Metabolic Insights and Process Optimisation

Nipon Sarmah[1], Tim Finnigan[2], Mark Taylor[3], Tom Vinestock[1], Ethan Errington[1], Miao Guo[1,*]

[1]Department of Engineering, Faculty of Natural, Mathematical & Engineering Sciences, King's College London, Strand Campus, London, WC2R 2LS, UK

[2]New Era Foods, Rudby Lea Hutton, Rudby, TS15 0JZ, UK

[3]Fermentation Lead, Marlow Ingredients, Nelson Ave, Billingham, North Yorkshire, TS23 4HA, UK

***Corresponding Author**
Miao Guo
E-mail address: miao.guo@kcl.ac.uk

**Abstract**

Fermentation-derived side streams represent an underutilised resource for sustainable protein production. This study investigates the potential of centrate from industrial *Fusarium venenatum* fermentation as a nutrient source for fungal biomass generation. Following compositional characterisation, a synthetic centrate medium was formulated and evaluated using a Box-Behnken design combined with response surface methodology. Across 46 experimental runs, cell dry weight (CDW) ranged from 0.22 to 3.87 g $L^{-1}$, demonstrating a strong dependence on nutrient composition. Ammonia and glucose were identified as the dominant factors influencing biomass production, with significant nonlinear effects. The model predicted a maximum CDW of 4.17 g $L^{-1}$ under optimised conditions, which was experimentally validated at 3.99 g $L^{-1}$. Carbon conversion efficiency reached up to 29.02%, indicating effective substrate utilisation. These findings demonstrate that fermentation-derived centrate can support substantial fungal growth, while highlighting its potential to enhance nutrient recovery and influence the biochemical composition of sustainable mycoprotein.

# 1. Introduction

Global demand for sustainable protein sources is increasing due to population growth, dietary shifts, and environmental constraints associated with conventional livestock production. Traditional animal agriculture is linked to high greenhouse gas emissions, extensive land use and significant freshwater consumption, prompting increasing interest in alternative protein sources that can be produced more efficiently and sustainably (Godfray et al., 2018; Poore & Nemecek, 2018). This challenge is further intensified by projections that the global population will exceed 10 billion by 2050 (Derbyshire et al., 2023).

Microbial protein production represents a promising strategy to address these challenges. Among microbial protein sources, mycoprotein derived from filamentous fungi has gained considerable commercial and scientific attention due to its high nutritional value, favourable amino acid composition and ability to be produced through controlled fermentation processes (Derbyshire et al., 2023; Finnigan et al., 2019). The filamentous fungus *Fusarium venenatum* A3/5 has been widely used for large-scale mycoprotein production since the 1980s and forms the basis of commercial products such as Quorn (Finnigan et al., 2017; Wiebe, 2002). Mycoprotein is characterised by a high protein content, dietary fibre, and a comparatively low environmental footprint relative to many animal-derived protein sources (Finnigan et al., 2019).

Despite the advantages of microbial protein production, industrial fermentation processes generate substantial quantities of side streams and residual process liquids. In industrial mycoprotein production, downstream processing generates a liquid fraction commonly referred to as centrate, which contains residual sugars, amino acids and salts (Finnigan et al., 2025). More broadly, fermentation and food-processing systems generate large volumes of nutrient-rich wastewater (~400 billion $m^3$ annually), characterised by high organic load and nutrient content, which increases both treatment requirements and opportunities for resource recovery (Durkin et al., 2024; Piercy et al., 2025).

In microbial protein systems, a significant fraction of carbon and biomass may remain in liquid side streams (up to ~30%), alongside dissolved sugars, polyols and metabolic intermediates (Banks et al., 2022; Piercy et al., 2025). Such streams are therefore increasingly recognised as potential resources for nutrient recovery and valorisation. At the same time, the agri-food sector generates large volumes of organic waste streams rich in recoverable carbon, highlighting the broader opportunity for circular bioprocessing (Piercy et al., 2022). However, these streams are often underutilised and may require additional treatment prior to disposal, representing both an environmental burden and a missed opportunity for nutrient recycling.

Previous studies have shown that fermentation wastewater streams contain soluble carbohydrates, nitrogen compounds and metabolites that can support microbial growth (Banks et al., 2022; Piercy et al., 2025). These streams are compositionally complex, containing residual substrates, volatile fatty acids and partially metabolised intermediates arising from incomplete substrate utilisation (Piercy et al., 2024). However, their composition can vary significantly throughout the fermentation cycle due to changes in substrate utilisation and metabolic activity (Wiebe, 2002). Such variability presents a key limitation for direct reuse, as fluctuations in nutrient availability can lead to inconsistent microbial performance and biomass yield, thereby necessitating controlled formulation or supplementation strategies for reliable application.

One approach to maximising the utilisation of such nutrient-rich streams is the optimisation of cultivation conditions using statistical experimental design techniques. Response surface methodology (RSM) is widely used in fermentation studies to evaluate the combined effects of multiple process variables and to determine optimal operating conditions with a reduced number of experimental trials (Montgomery, 2017). Among these, the Box–Behnken design (BBD) enables efficient estimation of quadratic and interaction effects while minimising the number of experimental runs (Bezerra et al., 2008; Sarmah et al., 2024).

While most studies have focused on fungal biomass production using defined media, relatively limited research has explored the utilisation of industrial fermentation side streams such as centrate (Ng et al., 2025; Sekoai et al., 2026). The valorisation of such streams offers an opportunity to enhance process sustainability by recovering residual nutrients and reintegrating them into production systems. However, effective integration requires controlled optimisation of nutrient composition rather than direct reuse, particularly to maintain consistent carbon-nitrogen balance and process performance. Improving utilisation of these streams could increase overall carbon efficiency while reducing downstream wastewater treatment burdens (Durkin et al., 2024; Piercy et al., 2022)

Therefore, the objective of this study was to evaluate the potential of fermentation-derived centrate as a nutrient source for fungal biomass production using *F. venenatum*, with particular emphasis on nutrient composition, metabolic utilisation, and process optimisation. A synthetic centrate medium was formulated based on compositional analysis to enable controlled experimentation, and a BBD combined with RSM was applied to identify optimal growth conditions. This study further aims to link nutrient availability with metabolic behaviour and biomass composition, providing insight into biochemical outcomes. This work contributes to the development of circular fermentation strategies for alternative protein production by demonstrating how industrial side streams can be valorised into nutritionally relevant microbial biomass through controlled and optimised utilisation. The overall experimental workflow employed in this study is summarised in Figure 1.

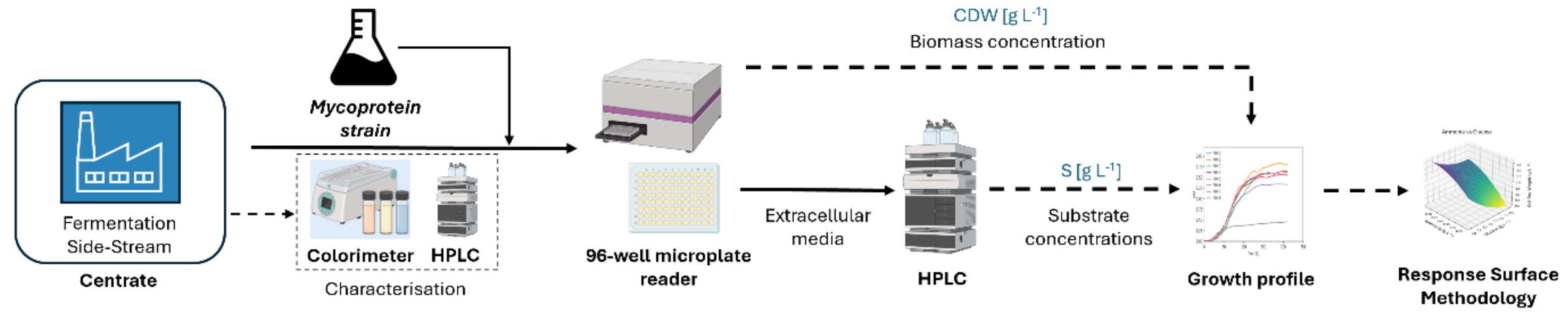

Figure 1: Experimental workflow for centrate valorisation and optimisation of fungal biomass production.

## 2. Materials and Methods

### 2.1 Microorganism and Inoculum Preparation

The filamentous fungus *F. venenatum* ATCC PTA 2684 was used as the production organism due to its established industrial application in mycoprotein production and its high protein yield and favourable amino acid profile. The strain was maintained on malt extract malt agar plates and stored at 4 °C for short-to medium-term preservation. The inoculum preparation procedure was adapted from previously reported protocols for *F. venenatum* cultivation (Banks et al., 2024). Prior to fermentation experiments, a preculture was prepared by transferring spores from the agar plate into a sterile 50 mL Erlenmeyer flask containing 10 mL nutrient medium supplemented with D-(+)-glucose (≥99%, Sigma Aldrich) (3 % w/v) and pH adjusted to 6.6. The flask was incubated in a shaking incubator (ES-20, Grant-Bio, UK) at 28 °C and 130 rpm for 96 h to allow sufficient fungal growth and sporulation.

Following incubation, the culture was filtered through a sterile 100 µm cell strainer to separate the conidia from the mycelial biomass, thereby ensuring a homogeneous spore suspension for reproducible inoculation. The concentration of conidia was determined using a Neubauer counting chamber (Hawksley), and the suspension was diluted with sterile nutrient medium to obtain a working inoculum concentration of approximately $10^3$ conidia $mL^{-1}$.

The prepared inoculum was subsequently used to initiate the fermentation experiments. All experimental procedures were conducted using sterile techniques within a Class II biological safety cabinet (ESCO, UK) to minimise the risk of contamination.

## 2.2 Centrate Source and Characterisation

Centrate used in this study was obtained from Marlow Ingredients (Billingham, United Kingdom), the industrial producer of Quorn mycoprotein derived from *F. venenatum*. Upon receipt, the centrate was stored at -20 °C to prevent compositional changes prior to analysis. The composition of the centrate was characterised to determine the concentrations of key nutrients relevant for fungal growth.

### *2.2.1 Determination of Ammonia Content*

The ammonia concentration of the centrate was determined using a Lovibond colorimetric analyser following the analytical protocol described by Piercy et al., 2025. A 2 mL aliquot of the centrate sample was added to a reaction vial containing cyanurate and salicylate reagents (Vario Ammonia Salicylate F5 Powder Pack and Vario Cyanurate F5 Powder Pack, Lovibond). The vials were securely closed and mixed until the reagents were completely dissolved. The reactions were incubated at room temperature for 20 min and measured at 600 nm using an MD600 spectrophotometer (Lovibond, Wiltshire, UK). All samples were analysed in triplicate to ensure analytical reproducibility. This method provides a quantitative estimate of bioavailable inorganic nitrogen present in the centrate.

### *2.2.2 Sugar Composition Analysis*

The carbohydrate composition of the centrate was determined using ultra-high-performance liquid chromatography (UHPLC) (Nexera L40, Shimadzu, Japan) equipped with SPD-M40 photodiode array and RID-20A refractive index detectors, following the analytical approach described by Banks et al., 2024 for quantifying fermentable sugars in fungal fermentation systems. Prior to analysis, samples were filtered through 0.22 µm membrane filters to remove suspended solids. For UHPLC analysis, 10 µL of the centrate sample was injected into an Aminex HPX-87H carbohydrate analysis column (Bio-Rad). The system was operated with 0.001 M sulfuric acid as the mobile phase at a flow rate of 0.4 mL $min^{-1}$, and a column temperature of 30 °C (Banks et al., 2024). Quantification of sugars was performed using external calibration curves generated from analytical standards. This approach enabled accurate identification and quantification of residual fermentable carbohydrates present in the centrate.

#### *2.2.3 Preparation of Synthetic Centrate*

To minimise variability associated with industrial waste streams and to ensure consistent experimental conditions, a synthetic centrate medium was prepared based on the compositional analysis of the original centrate. The synthetic formulation replicated the measured concentrations of ammonia and sugars identified during the characterisation stage. The synthetic centrate was subsequently used as the substrate for fermentation experiments, allowing systematic and controlled evaluation of nutrient effects independent of batch-to-batch variability inherent in industrial streams.

### 2.3 Experimental Design

The RSM approach was employed to investigate the effects of key process variables on fungal biomass production and to determine the optimal conditions for growth using synthetic centrate as the substrate. RSM is a statistical tool commonly used to evaluate the interactions between multiple variables and optimise bioprocess performance.

A BBD with five independent variables at three levels was used to construct the experimental matrix. The selected variables were ammonia, melibiose, glucose, mannitol, and arabitol, which were identified during the centrate characterisation stage as the principal nitrogen and carbon components present in the medium. Each factor was evaluated at three coded levels (−1, 0, +1) representing low, central, and high values, respectively. The experimental design consisted of 46 runs, including 6 replicated centre points, enabling estimation of experimental error and improving model reliability. All experiments were conducted in randomised order to minimise the effects of uncontrolled variables. The response variable measured in this study was cell dry weight (CDW, g $L^{-1}$), which was used to evaluate the growth performance of *F. venenatum* under the different experimental conditions.

## 2.4 Fermentation Experiments

A total of 46 fermentation experiments were conducted to evaluate the growth performance of *F. venenatum* using synthetic centrate as the substrate under the nutrient conditions defined by the BBD. The synthetic centrate medium was prepared according to the compositional ranges identified during the centrate characterisation stage, with nutrient concentrations adjusted according to the levels specified in the experimental design matrix. The medium was sterilised using 0.22 μm membrane filters prior to use to ensure aseptic conditions during cultivation. Fermentation cultures were initiated by inoculating the prepared medium with the previously prepared *F. venenatum* inoculum suspension.

Triplicate cultures for each experimental condition were dispensed into the wells of a transparent, sterile, flat-bottom, untreated 96-well microplate (VWR). The total working volume of each well was 150 μL with an initial inoculum concentration of $10^3$ conidia $mL^{-1}$. The outer wells of the microplate were filled with 150 μL ultra-pure water (Arium Pro, Sartorius, Germany) as blanks to minimise edge effects caused by evaporation. The lid of the 96-well plate was attached and sealed around the edges with parafilm to minimise evaporation. The plate was then placed in the tray of a microplate reader (FLUOstar Omega, BMG Labtech, Germany) and incubated at 28 °C under stationary conditions. Optical density (OD) at 600 nm was recorded from bottom-reading absorbance measurements taken at 2-h intervals, providing an indirect measurement of biomass growth in each well.

Growth curves were constructed by plotting $OD_{600}$ against time, enabling characterisation of different growth phases, e.g. lag, exponential, stationary and inhibitory phases. These profiles were used to assess fungal adaptation, growth kinetics, and the influence of nutrient composition on biomass accumulation.

At the end of the fermentation, the 96-well plate was removed from the microplate reader. The culture contents from each well were transferred into sterile 1.5 mL microcentrifuge tubes and centrifuged for 5 min at 1200 rpm. The supernatant was filtered through 0.2 µm regenerated cellulose syringe filters directly into 0.3 mL short-thread ND9 borosilicate glass vials for sugar analysis via UHPLC using the same analytical protocol described in Section 2.2.2.

To quantify biomass concentration at the end of the fermentation experiments, cell pellets from replicate wells were pooled to obtain a representative biomass sample. The pooled cell pellets were transferred onto pre-dried and pre-weighed Grade 698 glass fibre filters (VWR). The biomass was rinsed several times with ultrapure water under vacuum filtration to remove residual medium components. The filters containing the cell pellets were oven-dried at 80 °C for 24 h to obtain a constant weight. After drying, the filters were weighed using an analytical balance (Quintix Pro, Sartorius, Germany) to determine the dry biomass weight. The recorded mass was normalised by the initial culture volume to obtain the CDW (g $L^{-1}$).

Carbon conversion efficiency (CCE) was calculated by Equation (1) to evaluate the efficiency of carbon substrate utilisation for fungal biomass formation, based on the consumption of carbon-containing substrates, namely melibiose, glucose, mannitol, arabitol, and citric acid (from nutrient media), while ammonia was excluded as it does not contribute to the carbon balance.

$$Carbon\ conversion\ efficiency\ (\%) = \frac{X}{\sum_{i=1}^{n}(S_{i,0}-S_{i,f})} * 100 \quad (1)$$

where $X$ is the final biomass concentration (CDW, g $L^{-1}$), and $S_{i,0}$ and $S_{i,f}$ represent the initial and final concentrations of the $i^{th}$ carbon substrate, respectively.

## 2.5 Statistical Analysis

Statistical analysis of the experimental data was performed to evaluate the effects of the selected process variables on fungal biomass production and to determine the optimal cultivation conditions. The experimental data generated from the BBD were analysed using RSM, a widely applied statistical approach for modelling and optimisation of multivariable bioprocess systems.

The experimental data were fitted to a second-order (quadratic) polynomial regression model, incorporating linear, quadratic, and interaction terms. The statistical significance of the model and its coefficients was evaluated using analysis of variance (ANOVA). Model adequacy was assessed based on the coefficient of determination ($R^2$), adjusted $R^2$, and predicted $R^2$, which indicate the goodness-of-fit and predictive capability of the model.

The statistical significance of individual model terms, including linear, quadratic, and interaction effects, was determined using p-values at a 95% confidence level ($p < 0.05$). Terms with p-values less than 0.05 were considered statistically significant contributors to the model. In addition, the lack-of-fit test was used to evaluate whether the model adequately represented the experimental data, with a non-significant lack-of-fit indicating good model adequacy.

To further assess the relative importance of the process variables, Pareto charts of standardised effects were generated, allowing visual identification of the most influential factors affecting fungal biomass production. A significance threshold corresponding to $\alpha = 0.05$ was applied to identify statistically significant effects. In addition, three-dimensional response surface plots and two-dimensional contour plots were constructed to illustrate the interactions between variables and to identify optimal operating conditions within the experimental domain. All statistical analyses were performed using Minitab Statistical Software.

## 3. Results and Discussion

### 3.1 Chemical Characterisation of Centrate

Centrate generated during industrial mycoprotein production contained residual carbon substrates, soluble nitrogen species, and metabolic by-products that could support fungal growth, although its composition is expected to vary over the course of fermentation because of changing substrate consumption and metabolite formation (Finnigan et al., 2025; Wiebe, 2002). To preserve industrial relevance while reducing batch-to-batch variability, representative concentration ranges for the principal carbon and nitrogen components were established by integrating values reported in the literature (Piercy et al., 2025) with measurements obtained in the present study and used to formulate a synthetic centrate medium for controlled experimentation (Table 1).

The characterised centrate contained ammonia (0.02-0.325 g $L^{-1}$) together with melibiose, glucose, mannitol, and arabitol, indicating the presence of both assimilable nitrogen and mixed carbon substrates. Ammonia provides a readily available inorganic nitrogen source and is directly linked to amino acid and protein biosynthesis, making it a key determinant of fungal growth and biomass composition (Marzluf, 1997; Wiebe, 2002). Glucose is expected to be the preferred carbon source because it enters glycolysis directly and rapidly supplies ATP and precursor metabolites for biosynthesis (Ruijter & Visser, 2006). By contrast, melibiose requires prior hydrolysis by α-galactosidase before assimilation, which is likely to delay its contribution to biomass formation (Kabir et al., 2025). Mannitol and arabitol, as fungal polyols, are more likely to contribute to redox balancing and osmotic regulation than to rapid primary growth, consistent with their known physiological roles (Ruijter et al., 2003).

The coexistence of glucose, disaccharides, and polyols suggests mixed-substrate metabolism, but not equal substrate contribution. Carbon catabolite repression is likely to prioritise glucose utilisation and delay uptake of secondary substrates until simpler carbon sources are depleted (de Vries & Visser, 2001; Ruijter & Visser, 2006). This balance between carbon and nitrogen assimilation is important because it influences not only biomass yield but also the biochemical composition of the resulting fungal biomass, including protein content and cell wall carbohydrate fractions that contribute to nutritional and functional properties of mycoprotein (Wiebe, 2002). Overall, the centrate represented a nutrient-rich fermentation side stream suitable for fungal cultivation and justified its use as the basis for medium optimisation studies. Importantly, while the above compositional analysis provides a controlled basis for experimentation, it should be noted that the synthetic centrate formulation used in this study does not fully capture the compositional complexity of real industrial centrate streams. In practice, fermentation-derived centrate may contain additional components such as secondary metabolites, residual organic acids, and inorganic ions, which can influence microbial growth and metabolic activity (Lonchamp et al., 2022). In recirculating or partially closed-loop systems, the accumulation of such compounds over time may lead to inhibitory effects, osmotic stress, or shifts in metabolic flux distribution, thereby affecting process stability and biomass yield.

Furthermore, the concentration ranges used in the synthetic centrate were selected to represent typical values derived from compositional analysis and literature reports, rather than the full extent of temporal variability observed in industrial processes. Previous studies have shown that centrate composition can fluctuate significantly during fermentation due to dynamic substrate utilisation and metabolite formation (Lonchamp et al., 2022). Therefore, the results presented here should be interpreted within a controlled but industrially relevant compositional window, and future work should investigate the effects of real centrate variability and component accumulation under continuous or recirculating conditions.

### 3.2 Design of experiment and experimental data collection

A BBD was used to evaluate the effects of five centrate-derived variables, ammonia, melibiose, glucose, mannitol, and arabitol, on fungal biomass production. The design comprised 46 runs, including six replicated centre points, allowing efficient estimation of linear, quadratic, and interaction effects over the selected experimental space (Bezerra et al., 2008; Montgomery, 2017). The broad spread of response values confirmed that the chosen factor ranges successfully captured biologically meaningful variation in fungal performance (Table 2).

*3.2.1 Growth behaviour of Fusarium venenatum*

Growth curves for the 46 runs (Figure 2) showed the expected sigmoidal pattern of lag, exponential, and stationary phases, consistent with batch growth of filamentous fungi (Banks et al., 2024). The lag phase likely reflects physiological adaptation to the synthetic centrate medium, including activation of transport and enzymatic systems required for mixed-substrate assimilation as well as the morphological change from conidia to hyphae. This was followed by a clear exponential phase, during which rapid biomass accumulation was associated with the availability of readily assimilable carbon sources (primarily glucose), and then by a stationary phase associated with nutrient limitation, especially nitrogen depletion, or reduced availability of readily assimilable carbon.

Marked differences in growth behaviour were observed across the design space. Runs associated with higher final biomass showed steeper exponential phases and higher plateau OD values, whereas low-performing runs showed slower growth and lower final optical density. In particular, runs 23 and 24 achieved the highest CDW values (3.87 g $L^{-1}$), whereas runs 8, 15, 25, 29, 35, 38, and 39 showed markedly restricted growth, several with CDW below 1 g $L^{-1}$. Many of the poorest-performing conditions contained the lowest ammonia concentration (0.02 g $L^{-1}$), indicating that nitrogen limitation strongly constrained fungal proliferation even when carbon was present.

However, it is notable that final OD values did not always scale directly with CDW (e.g., runs 23 vs 24), suggesting that optical density alone may not fully capture biomass differences, potentially due to morphological variation or pellet formation in filamentous growth. These observations confirm that growth of *F. venenatum* in centrate-based media is governed primarily by nutrient balance rather than total substrate presence alone, with nitrogen availability emerging as a dominant limiting factor within the tested design space.

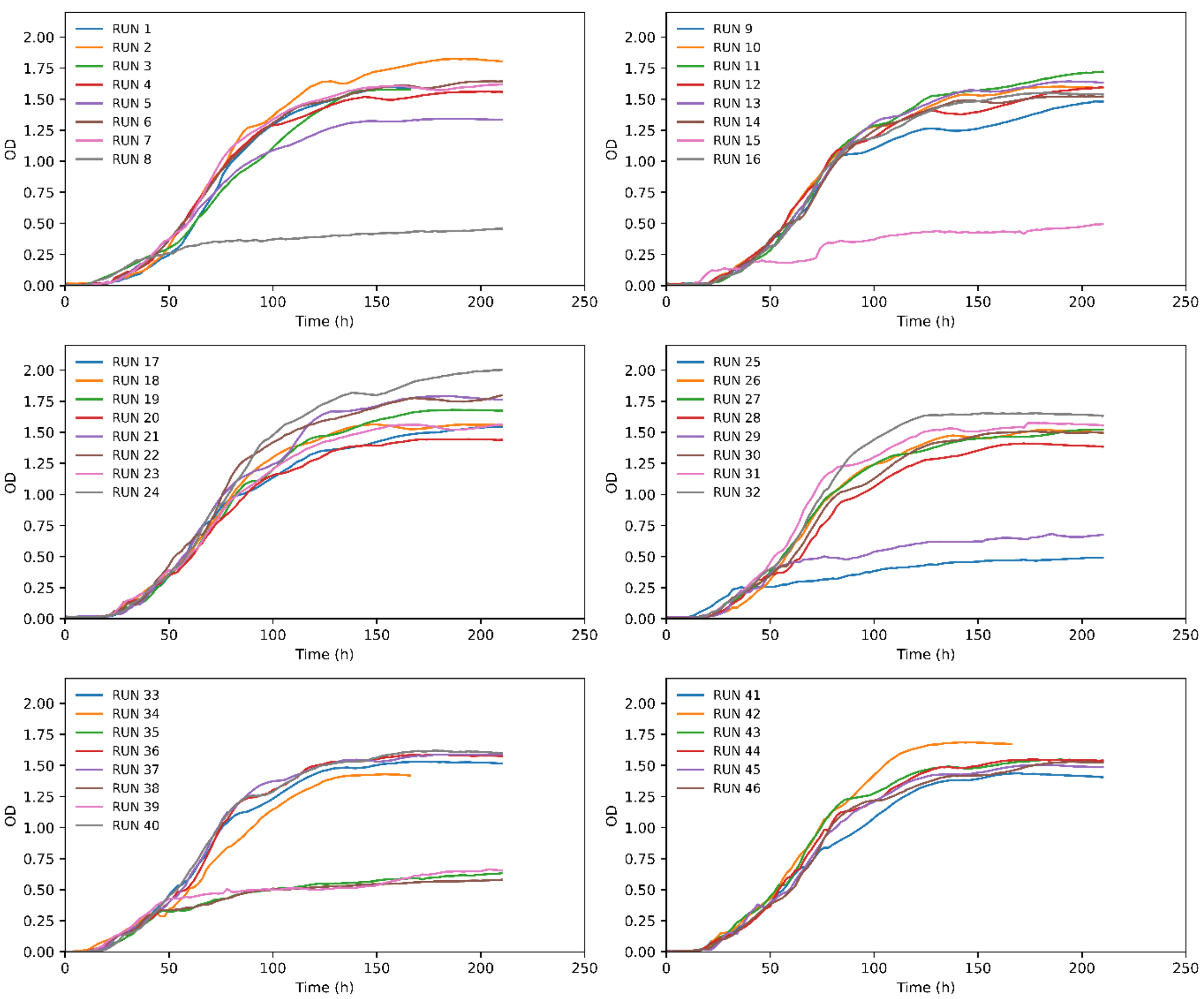

Figure 2: Growth curves of *Fusarium venenatum* across the 46 experimental runs conducted using the Box-Behnken design (BBD).

### *3.2.2 Biomass Production*

Consistent with the observed growth kinetics, final gravimetric measurements further confirmed substantial variation in fungal biomass production across the 46 experiments, with CDW values ranging from 0.22 to 3.87 g $L^{-1}$ as shown in Table 2, corresponding to an approximately 17.6-fold difference. Such a broad response range supports the suitability of the selected design space for response surface modelling and optimisation (Bezerra et al., 2008). Higher biomass concentrations were generally obtained under conditions combining sufficient ammonia with high glucose availability. The best-performing runs, particularly 23 and 24, were characterised by 4 g $L^{-1}$ glucose and moderate-to-high ammonia concentrations, indicating that these conditions supported efficient biomass synthesis. In contrast, low ammonia consistently suppressed growth, with runs 8, 15, 25, 29, 35, 38, and 39 all producing less than 1 g $L^{-1}$ CDW. This confirms nitrogen availability as a major limiting factor in centrate-based fungal cultivation, consistent with the central role of ammonium in amino acid and protein biosynthesis (Marzluf, 1997; Wiebe, 2002).

Glucose also exerted a strong positive effect. Conditions containing 4 g $L^{-1}$ glucose generally supported markedly higher CDW than those containing 0.5 g $L^{-1}$ glucose, highlighting the importance of a readily assimilable carbon source for ATP generation and precursor supply through glycolysis (Ruijter & Visser, 2006). The combined dominance of ammonia and glucose can be explained by their complementary metabolic roles: glucose supplies energy and carbon skeletons, while ammonium supports direct nitrogen incorporation via the GS-GOGAT pathway (Tesch et al., 1999). Together, these two variables govern the effective carbon-to-nitrogen balance of the medium and therefore the extent to which substrate carbon can be directed towards biomass rather than maintenance or overflow metabolism.

By comparison, melibiose, mannitol, and arabitol had weaker effects within the investigated range. Their more limited influence is consistent with their roles as secondary or slower-utilised carbon sources, which become important only after preferred substrates are depleted or under specific metabolic conditions (Ruijter et al., 2003; Ruijter & Visser, 2006). Biomass concentrations reported for *F. venenatum* vary widely depending on cultivation strategy and substrate availability. Laboratory-scale submerged fermentations using defined media typically achieve biomass concentrations of approximately 6.2 g $L^{-1}$ (Tong et al., 2024), while surface culture systems yield slightly lower values around 5.5 g $L^{-1}$ (Majumder et al., 2023). Laboratory-scale batch systems using defined or synthetic substrates can achieve biomass concentrations of approximately 10–11 g $L^{-1}$, as demonstrated in microlitre-scale fermentations on glucose–xylose mixtures (Banks et al., 2024). At an industrial scale, continuous fermentation processes have been reported to achieve biomass concentrations of ~15 g $L^{-1}$ (Wiebe, 2002). In this context, the lower biomass concentrations observed in the present study are consistent with cultivation on dilute, waste-derived substrates, where carbon availability and substrate quality inherently limit biomass accumulation relative to defined laboratory and industrial systems.

Importantly, the observed dependence of biomass production on carbon and nitrogen availability has implications beyond yield optimisation. Variations in nutrient composition are likely to influence the biochemical profile of the resulting fungal biomass, including protein content, amino acid distribution, and cell wall polysaccharide composition. Such compositional attributes are critical determinants of the nutritional quality and functional properties of mycoprotein-based foods (Wiebe, 2002). Therefore, optimisation of centrate-derived media should be considered not only in terms of biomass yield but also in relation to the compositional characteristics of the final product.

It is also important to consider that variations in carbon-to-nitrogen (C/N) ratio may influence not only biomass yield but also secondary metabolism in filamentous fungi. Nutrient imbalances, particularly nitrogen limitation in the presence of excess carbon, have been associated with the induction of secondary metabolite pathways, including mycotoxin production in certain fungal species (Brzonkalik et al., 2011). Although *F. venenatum* strains used in mycoprotein production are carefully selected and controlled to minimise such risks, the potential influence of centrate-derived nutrient variability on secondary metabolite formation warrants further investigation. Future work should therefore include targeted stress-testing under varying C/N conditions and analytical screening for mycotoxins to ensure process safety and robustness in food applications.

### *3.2.3 Substrate utilisation and Carbon conversion efficiency*

Substrate analysis showed that *F. venenatum* utilised multiple carbon sources present in the synthetic centrate medium. All substrates were almost completely consumed in the majority of runs, confirming effective utilisation of the available carbon pool. Glucose was consistently depleted, supporting its role as the primary and preferred carbon source. Mannitol and arabitol were also consumed to a substantial extent, demonstrating metabolic flexibility towards polyol utilisation. Citric acid, present at 7 g $L^{-1}$ in all experimental runs and melibiose were likewise utilised across most conditions, contributing additional carbon to the system. By contrast, residual glucose, mannitol, arabitol, melibiose, and citric acid were detected in low-performing runs, particularly 8, 15, 25, 29, 35, 38, and 39, indicating that poor growth was associated with incomplete substrate uptake.

This incomplete utilisation is most plausibly explained by nutrient imbalance, especially nitrogen limitation, which restricts overall metabolic activity and prevents efficient conversion of available carbon into biomass. Melibiose is also expected to be consumed more slowly than glucose because its assimilation depends on prior enzymatic hydrolysis, further reducing its immediate contribution to central metabolism.

CCE calculated from the utilisation of melibiose, glucose, mannitol, arabitol, and citric acid ranged from 5.83% to 29.02% (Table 2). The highest CCE was observed in run 24, which also produced one of the highest biomass concentrations, whereas the lowest CCE occurred in run 15, which showed poor growth. This positive association indicates that favourable nutrient conditions improve not only biomass accumulation but also the efficiency with which consumed carbon is channelled into fungal biomass.

CCE is particularly important from a sustainability perspective because it links substrate use to process efficiency, waste generation, and environmental burden (Poore & Nemecek, 2018). Higher CCE suggests more efficient anabolic use of carbon and potentially more favourable biomass composition, whereas lower values imply greater carbon loss to respiration or non-productive metabolism. In this study, the observed variability in CCE reinforces the conclusion that nutrient balance, especially the combined availability of glucose and ammonia, is central to efficient centrate valorisation.

These findings further emphasise that efficient carbon utilisation in mixed-substrate systems is closely linked to balanced nutrient availability, particularly the coordination between readily assimilable carbon and nitrogen sources. Such optimisation is essential not only for improving process efficiency but also for enhancing the sustainability profile of alternative protein production systems.

### 3.3 Model Fitting and Statistical Validation

RSM was applied to develop a predictive model describing the relationship between the selected process variables and fungal biomass production. A second-order polynomial regression model was fitted to the experimental data generated by the BBD to evaluate the combined effects of ammonia, melibiose, glucose, mannitol, and arabitol on the CDW of *F. venenatum*. Quadratic models are commonly used in bioprocess optimisation studies because they effectively capture linear, quadratic, and interaction effects between variables while requiring a relatively small number of experimental runs (Bezerra et al., 2008; Montgomery, 2017).

The fitted quadratic regression equation describing fungal biomass production in terms of the independent variables is presented in Equation (2).

$$CDW = 2.112 + 1.03\,A - 0.0031\,B + 0.867\,C + 0.694\,D + 0.058\,E - 0.446\,A^2 - 0.09\,B^2 + 0.395\,C^2 - 0.047\,D^2 - 0.084\,E^2 - 0.035\,AB - 0.255\,AC + 0.035\,AD - 0.135\,AE - 0.055\,BC + 0.055\,BD - 0.133\,BE + 0.247\,CD - 0.200\,CE + 0.055\,DE \quad (2)$$

where $A$, $B$, $C$, $D$, and $E$ represent the coded values of ammonia, melibiose, glucose, mannitol, and arabitol, respectively, and the coefficients correspond to the intercept, linear, quadratic, and interaction effects of the process variables on fungal biomass production.

The adequacy and statistical significance of the regression model were evaluated using ANOVA, and the results are summarised in Table 3. The ANOVA analysis demonstrated that the overall regression model was highly significant ($p < 0.001$), indicating that the selected variables collectively had a significant effect on fungal biomass production. The coefficient of determination for the model was $R^2$ = 94.22%, suggesting that the model explains the majority of the variability observed in the experimental data. Furthermore, the adjusted $R^2$ value of 89.60% indicates that the model remains robust after accounting for the number of predictors included in the regression equation. The predicted $R^2$ value of 80.05% further confirms the good predictive capability of the model for estimating biomass production within the experimental design space.

The low difference between the adjusted and predicted $R^2$ values indicate acceptable agreement between model fit and predictive performance and suggesting that the developed model was not overfitted. In addition, the lack-of-fit test was found to be statistically non-significant ($p = 0.600$), confirming that the regression model adequately described the experimental data and that the residual variation could be attributed primarily to random experimental error rather than systematic model inadequacy. Similar criteria are widely used in response surface optimisation studies to assess the validity and predictive reliability of quadratic models in bioprocess systems (Bezerra et al., 2008; Montgomery, 2017).

The relative influence of the process variables on fungal biomass production was further examined using a Pareto chart of standardised effects (Figure 3). Pareto analysis enables visual identification of the most influential model terms by ranking them according to their standardised effects relative to the significance threshold. At a significance level of $\alpha = 0.05$, the reference line corresponded to a standardised effect of 2.06. Based on the Pareto chart, the linear terms for ammonia (A) and glucose (C), together with the quadratic terms $A^2$ (AA) and $C^2$ (CC), exceeded this threshold and were therefore identified as statistically significant contributors to the model.

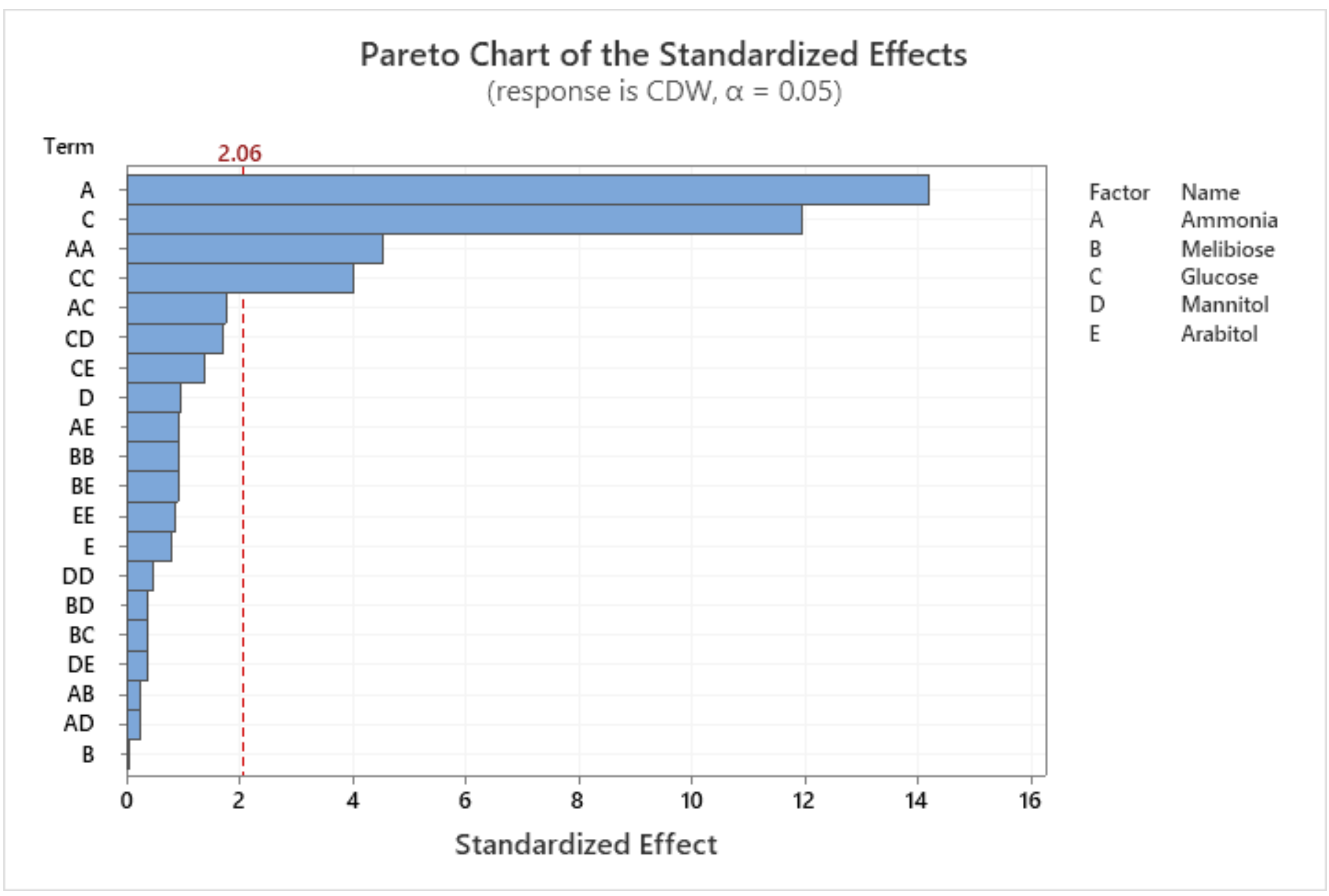

Figure 3: Pareto chart of standardised effects for the quadratic regression model describing fungal biomass production.

Among all terms, ammonia showed the largest standardised effect, followed by glucose, confirming that nitrogen availability and readily assimilable carbon were the dominant drivers of biomass production within the investigated design space. This observation is consistent with the biological roles of these nutrients, as ammonia directly supports amino acid and protein synthesis, while glucose serves as a rapidly metabolised carbon source that provides both energy and precursor metabolites for biomass formation (Cairns et al., 2018; Wiebe, 2002).

The single-factor response curve for ammonia in Figure 4 showed the expected concave shape. This curvature was supported by the statistical significance of the quadratic ammonia term (Figure 3). When nitrogen is limiting, the biomass was approximately proportional to the initial ammonia concentration. However, when surplus nitrogen is available, this relationship breaks down as the carbon substrates become limiting. The predicted biomass at the lowest ammonia concentration slightly exceeded the stoichiometric expectation, which may reflect experimental uncertainty in the biomass measurement or limitations of the fitted quadratic model at the edge of the design space. The stoichiometric limit was calculated based on a molecular formula for *F. venenatum* of $CH_{1.98}N_{0.173}O_{0.647}P_{0.0151}$ (Upcraft et al., 2021). The quadratic RSM model should also be treated with caution when extrapolating; while the quadratic Equation for ammonia predicts symmetry about the maximum CDW achieved at approximately 0.35 g $L^{-1}$ initial ammonia, in practice, the drop-off in CDW may not be achieved until significantly higher ammonia concentrations begin to cause inhibition.

By contrast, the single-factor response curve for glucose in Figure 4 is convex, which was an unexpected result. The quadratic glucose term responsible for this curvature is also statistically significant (Figure 3). This could suggest that high initial glucose concentrations allow for greater utilisation of secondary substrates, such as citric acid, or even that the glucose yield is not constant but increases as the initial concentration of glucose increases. However, this interpretation should be treated with caution, as the observed convexity could also be due to experimental error, with the true underlying response being closer to linear or weakly concave.

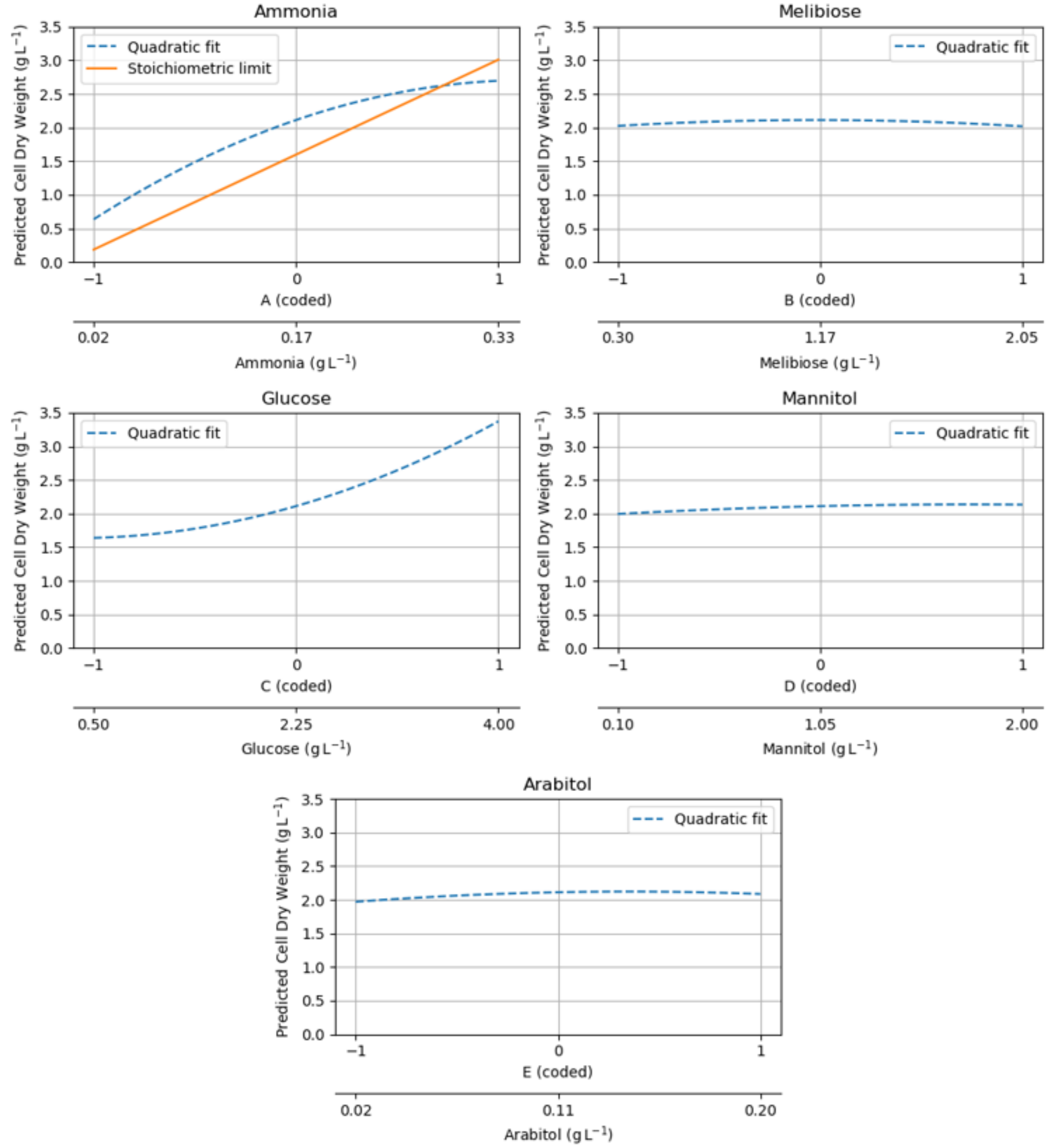

Figure 4: One-factor response curves showing the effect of individual variables on predicted cell dry weight (CDW) of *Fusarium venenatum*, with all other variables held at their central levels.

In contrast, the effects of melibiose (B), mannitol (D), and arabitol (E) were comparatively less pronounced within the investigated experimental domain. Their linear terms did not exceed the significance threshold in the Pareto chart, indicating that their individual contributions to CDW were relatively minor compared with those of ammonia and glucose. Similarly, Figure 4 shows that the effects of these variables on predicted CDW are minimal. This behaviour can be explained by the metabolic hierarchy of carbon utilisation in filamentous fungi, where simple sugars such as glucose are preferentially metabolised over more complex carbohydrates and polyols through carbon catabolite repression and related regulatory mechanisms (Ruijter & Visser, 2006). Although melibiose, mannitol, and arabitol were utilised, it is unclear from these results whether these substrates meaningfully contribute to the biomass production. In the case of arabitol, this may simply reflect the low concentrations tested. However, there is little evidence that these sugars caused inhibition at the concentrations tested.

Several interaction terms, particularly AC, CD, and CE, showed moderate standardised effects but remained below the significance threshold, suggesting that although interactions between variables may have influenced biomass formation to some extent, their contributions were weaker than the dominant main and quadratic effects of ammonia and glucose. This finding is consistent with the earlier experimental observations, where biomass production responded most strongly to nitrogen availability and glucose concentration, while the effects of other substrates were comparatively secondary.

Overall, the statistical analysis confirms that the developed quadratic regression model provides an accurate and reliable representation of fungal biomass production within the experimental design space. The high coefficient of determination, strong agreement between adjusted and predicted $R^2$ values, significant model terms, and a non-significant lack-of-fit test collectively demonstrate the suitability of the model for describing the effects of nutrient composition on fungal growth. The Pareto analysis further clarified that ammonia and glucose, together with their quadratic effects, were the principal determinants of biomass production. The validated model was subsequently used to generate response surface plots and to identify optimal nutrient conditions for maximising fungal biomass production.

### 3.4 Response Surface and Contour Plot Analysis

Response surface and contour plot analysis (Figure 5) provided further insight into the combined effects of nutrient variables on fungal biomass production. The steepest contour gradients were consistently associated with ammonia and glucose, whereas surfaces involving melibiose, mannitol, and arabitol were flatter, indicating weaker effects within the investigated range. This visual pattern supports the ANOVA and Pareto results and confirms that ammonia and glucose dominate the response landscape.

The plots demonstrated that ammonia concentration exerted a strong influence on fungal biomass production. Increasing ammonia from 0.02 to 0.325 g $L^{-1}$ increased predicted CDW across multiple variable combinations, demonstrating that nitrogen availability was a critical determinant of growth. Likewise, increasing glucose from 0.5 to 4 g $L^{-1}$ generally increased predicted biomass, especially when ammonia was not limiting. The ammonia-glucose interaction produced the most favourable response region, with the highest predicted CDW occurring at moderate-to-high ammonia and high glucose concentrations, consistent with the experimental performance of runs 23 and 24. This region represents the most favourable carbon-to-nitrogen balance for biomass formation within the design space.

In contrast, changes in melibiose, mannitol, and arabitol concentration caused only modest shifts in predicted CDW, indicating that their individual effects were comparatively small. This is biologically plausible because melibiose requires prior hydrolysis before entering central metabolism, while mannitol and arabitol are more closely linked to secondary carbon metabolism and redox homeostasis than to rapid biomass accumulation (Ruijter et al., 2003; Ruijter & Visser, 2006). Overall, the response surface analysis confirms that nutrient optimisation in centrate-based media should focus primarily on accessible nitrogen and readily metabolised carbon rather than on secondary substrates.

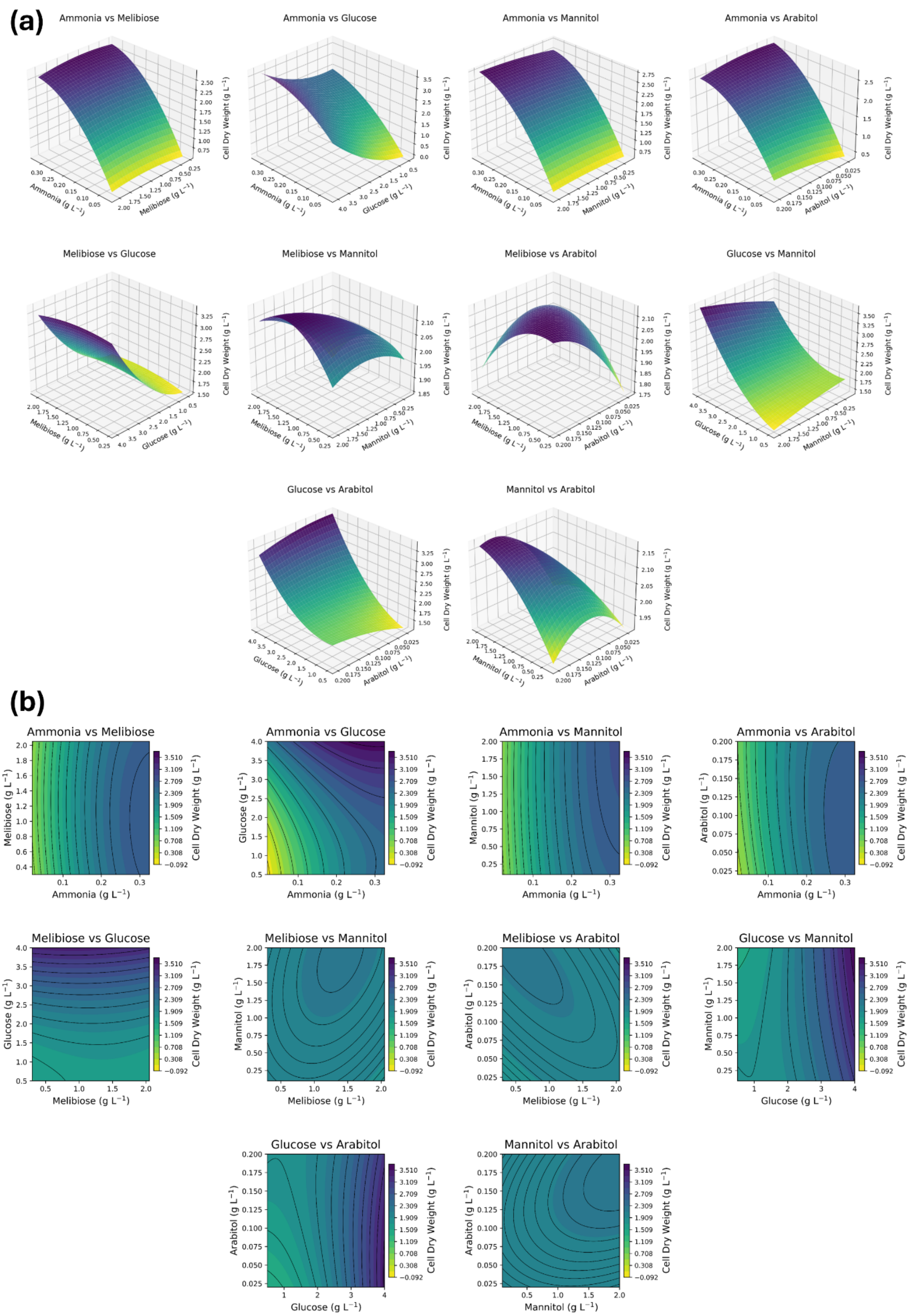

Figure 5: (a) Response surface and (b) contour plots showing the effects of nutrient

variables on fungal biomass production.

## 3.5 Process Optimisation

Building on the validated regression model and response surface analysis, numerical optimisation was performed using the desirability function approach to determine the optimal nutrient composition for maximising fungal biomass production. Numerical optimisation predicted that maximum biomass production would be achieved at 0.325 g $L^{-1}$ ammonia, 1.63 g $L^{-1}$ melibiose, 4.0 g $L^{-1}$ glucose, 2.0 g $L^{-1}$ mannitol, and 0.02 g $L^{-1}$ arabitol in addition to the 7 g $L^{-1}$ of citric acid used as a standard across all experiments, corresponding to a predicted CDW of 4.17 g $L^{-1}$. Validation experiments under these conditions produced 3.99 g $L^{-1}$ CDW, representing a deviation of only 4.3%, and confirming the predictive reliability of the model. CCE under the optimised condition was 27.24%, indicating effective use of the available carbon pool for biomass formation.

When compared with a higher-input, glucose-based defined medium containing 30 g $L^{-1}$ glucose and formulated at an overall C:N molar ratio of approximately 15:1, the centrate-based system produced a lower biomass concentration (3.99 vs 15.05 g $L^{-1}$) and lower CCE (27.24% vs 41.58%). However, this comparison must be interpreted in light of substrate input. The defined medium relied on substantially higher concentrations of refined substrates, whereas the centrate medium achieved meaningful biomass production using a lower-input, waste-derived nutrient mixture. This highlights a key trade-off between absolute biomass yield and resource efficiency: the defined medium maximises production, whereas the centrate system improves circularity, substrate recovery, and waste valorisation.

The optimisation results also reinforce the earlier statistical interpretation, as the predicted optimum lay at the upper tested bounds for ammonia and glucose, again confirming their dominant role in fungal growth. Taken together, the validation data show that centrate-derived media can support substantial biomass formation with reasonable conversion efficiency, even though they do not match the productivity of highly enriched defined media.

### 3.6 Implications for centrate valorisation and circular fermentation

Taken together, the results of this study demonstrate that fermentation-derived centrate should not be viewed solely as a waste stream requiring treatment, but as a recoverable nutrient source that can be strategically reintegrated into microbial protein production through controlled formulation and optimisation. The successful cultivation of *F. venenatum* on a synthetic centrate formulation, combined with optimisation to nearly 4 g $L^{-1}$ biomass, shows that fermentation side-streams can support meaningful fungal growth without exclusive reliance on refined inputs.

An important advantage of centrate is that it contains a mixture of carbon substrates, polyols, and nitrogen species that can be exploited by metabolically versatile filamentous fungi. This reduces the need for extensive feedstock standardisation and supports the use of heterogeneous industrial side-streams in secondary cultivation processes (Wiebe, 2002). From a circular bioeconomy perspective, the controlled reutilisation of such streams has a dual benefit: it lowers the burden of effluent treatment and simultaneously generates additional value as fungal biomass (Finnigan et al., 2019; Poore & Nemecek, 2018).

For mycoprotein production specifically, centrate valorisation offers a route to improve nutrient recovery, reduce waste generation, and enhance overall process sustainability. Although centrate-based media produced lower absolute biomass than defined media, they achieved this with much lower refined substrate input, which shifts the optimisation target from yield maximisation alone to resource-efficient production. This is particularly relevant in the alternative protein sector, where economic scalability must increasingly be aligned with environmental performance.

Overall, the study establishes centrate as a promising feedstock for circular fungal bioprocessing and provides a framework for its optimisation through RSM. However, in practical applications, centrate is unlikely to be reused directly due to compositional variability and potential accumulation of inhibitory compounds, and is more appropriately applied as a partially substituted or supplemented nutrient stream within a controlled bioprocessing system. Future work should focus on validation with real, non-synthetic centrate, scale-up under controlled bioreactor conditions, techno-economic and life cycle assessment to quantify industrial feasibility and environmental benefit, and comprehensive safety assessment of the resulting mycoprotein, including evaluation of protein quality, potential allergenicity, and the presence of undesirable metabolites or mycotoxins, which were not assessed in the present study. The absence of detailed compositional and safety profiling represents a limitation of this work and should be addressed to ensure suitability for food applications. Furthermore, untargeted metabolomic profiling is recommended to characterise the full metabolic landscape under centrate-based cultivation, enabling identification of pathway regulation and detection of any safety-relevant compounds.

## 4. Conclusion

This study demonstrates the potential of fermentation-derived centrate as a viable substrate for fungal biomass production using *F. venenatum*. Compositional analysis confirmed the presence of assimilable carbon sources and inorganic nitrogen, supporting its suitability as a growth medium. Through the application of RSM, nutrient composition was systematically optimised, with ammonia and glucose identified as the dominant factors governing biomass production. The developed model showed strong predictive capability, and experimental validation confirmed a maximum biomass concentration of 3.99 g $L^{-1}$ under optimised conditions.

Although biomass yields were lower than those obtained with conventional defined media, the centrate-based system achieved meaningful biomass production with substantially lower refined substrate input, demonstrating improved resource efficiency. CCE of up to 29.02% further indicate effective utilisation of mixed carbon substrates under favourable conditions.

Importantly, the results highlight that nutrient composition not only controls biomass yield but is also likely to influence the biochemical characteristics of the resulting mycoprotein, including protein content and structural components relevant to food functionality. This reinforces the importance of integrating metabolic and compositional considerations into process optimisation.

Overall, this work establishes fermentation-derived centrate as a promising feedstock for circular bioprocessing and provides a framework for its optimisation. In practical applications, centrate is best utilised through controlled formulation or supplementation rather than direct reuse, to ensure consistent nutrient balance and process performance. The integration of centrate valorisation into mycoprotein production systems offers a viable pathway to enhance sustainability, improve nutrient recovery, and reduce reliance on refined substrates in alternative protein production.

**Declaration of competing interests**

The authors declare the following financial interests/personal relationships which may be considered as potential competing interests:

Nipon Sarmah reports financial support was provided by The Good Food Institute Inc. Miao Guo reports financial support was provided by The Good Food Institute Inc. If there are other authors, they declare that they have no known competing financial interests or personal relationships that could have appeared to influence the work reported in this paper.

## Acknowledgements

MG and NS would like to acknowledge the generous support from the Good Food Institute (GFI) who kindly funded our research project "TRANSFORM: Waste-to-Microbial Protein" under the GFI Field Catalyst Grant Program.

Table 1: Experimental ranges and coded levels of the independent variables used in the Box-Behnken design (BBD) for optimisation of fungal biomass production using synthetic centrate medium.

| **Media components** | **Range** | **Level** | | |
|---|---|---|---|---|
| | | -1 | 0 | +1 |
| Ammonia (g/L) | 0.02 - 0.325 | 0.02 | 0.1725 | 0.325 |
| Melibiose (g/L) | 0.3 - 2.05 | 0.3 | 1.175 | 2.05 |
| Glucose (g/L) | 0.5 - 4 | 0.5 | 2.25 | 4 |
| Mannitol (g/L) | 0.1 - 2 | 0.1 | 0.5 | 2 |
| Arabitol (g/L) | 0.02 - 0.2 | 0.02 | 0.11 | 0.2 |

Table 2: Box-Behnken experimental design matrix and corresponding response values for fungal biomass production and carbon conversion efficiency (CCE).

| Run | Ammonia (g/L) | Melibiose (g/L) | Glucose (g/L) | Mannitol (g/L) | Arabitol (g/L) | Cell Dry Weight (g/L) | Carbon Conversion Efficiency (%) |
|---|---|---|---|---|---|---|---|
| 1 | 0.1725 | 1.175 | 4 | 1.05 | 0.02 | 2.8 | 21.140 |
| 2 | 0.325 | 1.175 | 2.25 | 1.05 | 0.2 | 2.53 | 21.670 |
| 3 | 0.325 | 2.05 | 2.25 | 1.05 | 0.11 | 2.53 | 20.305 |
| 4 | 0.1725 | 1.175 | 2.25 | 1.05 | 0.11 | 1.56 | 13.466 |
| 5 | 0.1725 | 1.175 | 0.5 | 1.05 | 0.2 | 2 | 20.151 |
| 6 | 0.1725 | 1.175 | 4 | 1.05 | 0.2 | 2.89 | 21.527 |
| 7 | 0.1725 | 2.05 | 4 | 1.05 | 0.11 | 3.11 | 21.886 |
| 8 | 0.02 | 0.3 | 2.25 | 1.05 | 0.11 | 0.27 | 7.634 |
| 9 | 0.1725 | 1.175 | 0.5 | 1.05 | 0.02 | 1.11 | 11.390 |
| 10 | 0.1725 | 1.175 | 4 | 0.1 | 0.11 | 3.11 | 25.111 |
| 11 | 0.1725 | 2.05 | 2.25 | 0.1 | 0.11 | 1.78 | 15.465 |
| 12 | 0.1725 | 0.3 | 2.25 | 1.05 | 0.02 | 2 | 18.832 |
| 13 | 0.1725 | 1.175 | 2.25 | 2 | 0.02 | 2 | 16.071 |
| 14 | 0.1725 | 0.3 | 2.25 | 2 | 0.11 | 2 | 17.153 |
| 15 | 0.02 | 2.05 | 2.25 | 1.05 | 0.11 | 0.27 | 5.832 |
| 16 | 0.1725 | 1.175 | 2.25 | 0.1 | 0.2 | 2 | 18.648 |
| 17 | 0.1725 | 1.175 | 2.25 | 0.1 | 0.02 | 2 | 18.966 |
| 18 | 0.1725 | 0.3 | 4 | 1.05 | 0.11 | 3.33 | 26.726 |
| 19 | 0.325 | 1.175 | 0.5 | 1.05 | 0.11 | 2.44 | 24.809 |
| 20 | 0.1725 | 1.175 | 0.5 | 2 | 0.11 | 1.33 | 12.332 |
| 21 | 0.325 | 1.175 | 2.25 | 0.1 | 0.11 | 2.53 | 23.789 |
| 22 | 0.1725 | 2.05 | 2.25 | 1.05 | 0.2 | 2 | 16.735 |
| 23 | 0.1725 | 1.175 | 4 | 2 | 0.11 | 3.87 | 27.091 |
| 24 | 0.325 | 1.175 | 4 | 1.05 | 0.11 | 3.87 | 29.021 |
| 25 | 0.02 | 1.175 | 2.25 | 1.05 | 0.02 | 0.27 | 6.922 |
| 26 | 0.1725 | 1.175 | 2.25 | 1.05 | 0.11 | 2 | 17.264 |
| 27 | 0.1725 | 2.05 | 2.25 | 2 | 0.11 | 2 | 14.914 |
| 28 | 0.1725 | 0.3 | 0.5 | 1.05 | 0.11 | 1.56 | 17.411 |
| 29 | 0.02 | 1.175 | 2.25 | 2 | 0.11 | 0.53 | 12.557 |
| 30 | 0.1725 | 0.3 | 2.25 | 0.1 | 0.11 | 2 | 20.492 |
| 31 | 0.1725 | 1.175 | 2.25 | 1.05 | 0.11 | 2.22 | 19.163 |
| 32 | 0.325 | 1.175 | 2.25 | 2 | 0.11 | 2.67 | 21.300 |
| 33 | 0.1725 | 2.05 | 0.5 | 1.05 | 0.11 | 1.56 | 14.566 |
| 34 | 0.325 | 1.175 | 2.25 | 1.05 | 0.02 | 2.67 | 23.227 |
| 35 | 0.02 | 1.175 | 2.25 | 0.1 | 0.11 | 0.53 | 13.534 |
| 36 | 0.1725 | 1.175 | 2.25 | 1.05 | 0.11 | 2.22 | 19.163 |
| 37 | 0.1725 | 1.175 | 2.25 | 2 | 0.2 | 2.22 | 17.584 |
| 38 | 0.02 | 1.175 | 0.5 | 1.05 | 0.11 | 0.22 | 6.721 |
| 39 | 0.02 | 1.175 | 2.25 | 1.05 | 0.2 | 0.67 | 12.231 |
| 40 | 0.1725 | 1.175 | 2.25 | 1.05 | 0.11 | 2.27 | 19.594 |
| 41 | 0.1725 | 1.175 | 0.5 | 0.1 | 0.11 | 1.56 | 17.558 |
| 42 | 0.325 | 0.3 | 2.25 | 1.05 | 0.11 | 2.67 | 24.930 |
| 43 | 0.1725 | 1.175 | 2.25 | 1.05 | 0.11 | 2.4 | 20.716 |
| 44 | 0.02 | 1.175 | 4 | 1.05 | 0.11 | 2.67 | 20.022 |
| 45 | 0.1725 | 0.3 | 2.25 | 1.05 | 0.2 | 2 | 18.519 |
| 46 | 0.1725 | 2.05 | 2.25 | 1.05 | 0.02 | 2.53 | 20.453 |

Table 3: Analysis of variance (ANOVA) for the quadratic regression model describing the effects of nutrient composition on fungal biomass production

| Source | DF | Adj SS | Adj MS | F-Value | P-Value |
|---|---|---|---|---|---|
| Model | 20 | 34.3338 | 1.7167 | 20.39 | 0.000 |
| *Linear* | 5 | 29.1292 | 5.8258 | 69.20 | 0.000 |
| Ammonia | 1 | 16.9744 | 16.9744 | 201.62 | 0.000 |
| Melibiose | 1 | 0.0002 | 0.0002 | 0.00 | 0.966 |
| Glucose | 1 | 12.0236 | 12.0236 | 142.81 | 0.000 |
| Mannitol | 1 | 0.0770 | 0.0770 | 0.91 | 0.348 |
| Arabitol | 1 | 0.0541 | 0.0541 | 0.64 | 0.431 |
| *Square* | 5 | 4.3503 | 0.8701 | 10.33 | 0.000 |
| Ammonia*Ammonia | 1 | 1.7379 | 1.7379 | 20.64 | 0.000 |
| Melibiose*Melibiose | 1 | 0.0713 | 0.0713 | 0.85 | 0.366 |
| Glucose*Glucose | 1 | 1.3588 | 1.3588 | 16.14 | 0.000 |
| Mannitol*Mannitol | 1 | 0.0193 | 0.0193 | 0.23 | 0.636 |
| Arabitol*Arabitol | 1 | 0.0612 | 0.0612 | 0.73 | 0.402 |
| *2-Way Interaction* | 10 | 0.8543 | 0.0854 | 1.01 | 0.459 |
| Ammonia*Melibiose | 1 | 0.0049 | 0.0049 | 0.06 | 0.811 |
| Ammonia*Glucose | 1 | 0.2601 | 0.2601 | 3.09 | 0.091 |
| Ammonia*Mannitol | 1 | 0.0049 | 0.0049 | 0.06 | 0.811 |
| Ammonia*Arabitol | 1 | 0.0729 | 0.0729 | 0.87 | 0.361 |
| Melibiose*Glucose | 1 | 0.0121 | 0.0121 | 0.14 | 0.708 |
| Melibiose*Mannitol | 1 | 0.0121 | 0.0121 | 0.14 | 0.708 |
| Melibiose*Arabitol | 1 | 0.0702 | 0.0702 | 0.83 | 0.370 |
| Glucose*Mannitol | 1 | 0.2450 | 0.2450 | 2.91 | 0.100 |
| Glucose*Arabitol | 1 | 0.1600 | 0.1600 | 1.90 | 0.180 |
| Mannitol*Arabitol | 1 | 0.0121 | 0.0121 | 0.14 | 0.708 |
| Error | 25 | 2.1048 | 0.0842 | | |
| Lack-of-Fit | 20 | 1.6563 | 0.0828 | 0.92 | 0.600 |
| Pure Error | 5 | 0.4485 | 0.0897 | | |
| Total | 45 | 36.4386 | | | |